\documentclass[runningheads]{llncs}

\usepackage{graphicx}
\usepackage{amsmath}
\usepackage{amssymb}
\usepackage{wrapfig}

\usepackage{multirow}
\usepackage[ruled,vlined]{algorithm2e}
\usepackage[title]{appendix}

\begin{document}

\title{Triple-View Feature Learning\\for Medical Image Segmentation}

% \author{Ziyang Wang \and Irina Voiculescu}
% \institute{Department of Computer Science, University of Oxford, UK \\
% \email{ziyang.wang@cs.ox.ac.uk}}

\author{Ziyang Wang \and Irina Voiculescu}
\institute{Department of Computer Science, University of Oxford, UK \\
\email{ziyang.wang@cs.ox.ac.uk}
}
\maketitle

%%%%%%%%% ABSTRACT
\begin{abstract}

Deep learning models, e.g.\ supervised Encoder-Decoder style networks, exhibit promising performance in medical image segmentation, but come with a high labelling cost. We propose TriSegNet, a semi-supervised semantic segmentation framework. It uses {\em triple-view\/} feature learning on a limited amount of labelled data and a large amount of unlabeled data. The triple-view architecture consists of three pixel-level classifiers and a low-level shared-weight learning module. The model is first initialized with labelled data. Label processing, including data perturbation, confidence label voting and unconfident label detection for annotation, enables the model to train on labelled and unlabeled data simultaneously. The confidence of each model gets improved through the other two views of the feature learning. This process is repeated until each model reaches the same confidence level as its counterparts. This strategy enables triple-view learning of generic medical image datasets. Bespoke overlap-based and boundary-based loss functions are tailored to the different stages of the training. The segmentation results are evaluated on four publicly available benchmark datasets including Ultrasound, CT, MRI, and Histology images. Repeated experiments demonstrate the effectiveness of the proposed network compared against other semi-supervised algorithms, across a large set of evaluation measures.

\end{abstract}

%%%%%%%%% BODY TEXT
\section{Introduction}

%The segmentation of organs or other regions of interest is often an essential prerequisite for computer-aided diagnosis, robotic surgery and medical image analysis.
The promising performance of deep learning for medical imaging relies not only on network architecture engineering, but also on the availability of sufficient high-quality manually annotated data, which is hard to come by.
% To tackle the challenge of high-cost labelling and limit resources of data in the medical imaging community, self-supervised learning, weakly supervised learning, and semi-supervised learning are widely studied. { \bf Self-supervised} learning is to learn visual features by proposing various pretext tasks, then learn large-scale unlabeled images in a supervised manner with transfer learning~\cite{jing2020self}. To reduce the high cost of labeling data, { \bf weakly-supervised} learning is studied that allows model learn features with low-quality labels such as bounding boxes, scribbles, and tags. 
Co-training, and self-training are two widely studied approaches in semi-supervised learning. 

%  The performance of weakly supervised learning and self-supervised learning, however, in medical image segmentation community is still limited. 

{ \bf Self-training} first initializes a model with labelled data. Then the model generates pseudo masks for unlabelled data. A condition is set for the selection of pseudo masks, and the model is retrained by expanding its training data~\cite{you2011segmentation}. { \bf Co-training} is normally used to train two separate models with two views which  benefit each other~\cite{blum1998combining} by expanding the size of the training data. Deep co-training was first proposed by~\cite{qiao2018deep} pointing out the challenge of `collapsed neural networks': training two models on the same dataset cannot enable multi-view feature learning because the two models will necessarily end up similar. %`View Difference Constraint' and a loss function were then proposed. 
%Disagreement-based multi-view Tri-net then proposed 
Output smearing, diversity augmentation, and dropout for pseudo label editing were designed to mitigate that effect~\cite{dong2018tri}.

%Besides encouraging the differences of the multi views, an uncertainty-aware scheme and other techniques aiming to generate and utilize reliable pseudo labels have also been studied~\cite{xia20203d}. 
%Current key studies of co-training rely on enabling the diversity of multi views, and on confidently generating pseudo labels. 

\section{Methodology}

%We propose three different CNN architectures of pixel-level classifiers as multi-view co-training framework with tailored data perturbation, pseudo label processing, multi-training stage strategy, and dual-stage loss design in this paper.

The central idea in our method is that several (three in our case) different views of the data and their associated learned parameters are developed in separate models simultaneously. This gives each  model a chance to complement the  others in what they contribute to the learning process. Moreover, they do not each process the same dataset: rather, they take carefully crafted disjoint parts of the data and learn what they can from each of those subsets.

%Whilst there have been other models which combine learning outcomes from three views of a dataset, ours is radically different from those in a number of distinct ways, discussed below.

A multi-view co-training semi-supervised learning method for classification is also proposed in \cite{dong2018tri}, but for a classification task, not for semantic segmentation.  Detailed mask prediction requires an entirely novel technique \cite{chen2021semi,wang2022icip}. 
Our pseudo-labels are generated afresh at intermediate steps selecting from amongst the output of two individual CNN components in order to train the third one. Our confidence estimation is  different from the uncertainty-aware scheme~\cite{xia20203d,wang2022uncertainty}: the triple framework decides which pseudo-labels to propagate further through selecting those in which the individual component models have had high confidence and have `voted' for. This also has the effect of increasing the overall confidence of the framework in a way that we specifically show how to measure.  Pseudo-labels with too low a confidence level are not used in the following step. How low that level of confidence depends on the stage of the training: the further into the training process, the more confident the framework needs to become. Thus the number of images available for training gets increased gradually.

%Cross Pseudo Supervision (\rev{2} \& \cite{chen2021semi}) proposes a pseudo-label-based method for a semi-supervised learning task. They follow the classical pattern of two CNNs with the same architecture learning the same feature information input, although their two sets of parameters are initialised separately. They outperform~\cite{ke2020guided}, \cite{tarvainen2017mean}, \cite{zou2020pseudoseg}.

\begin{figure*}[t]
% \centering  
\includegraphics[width=\linewidth]{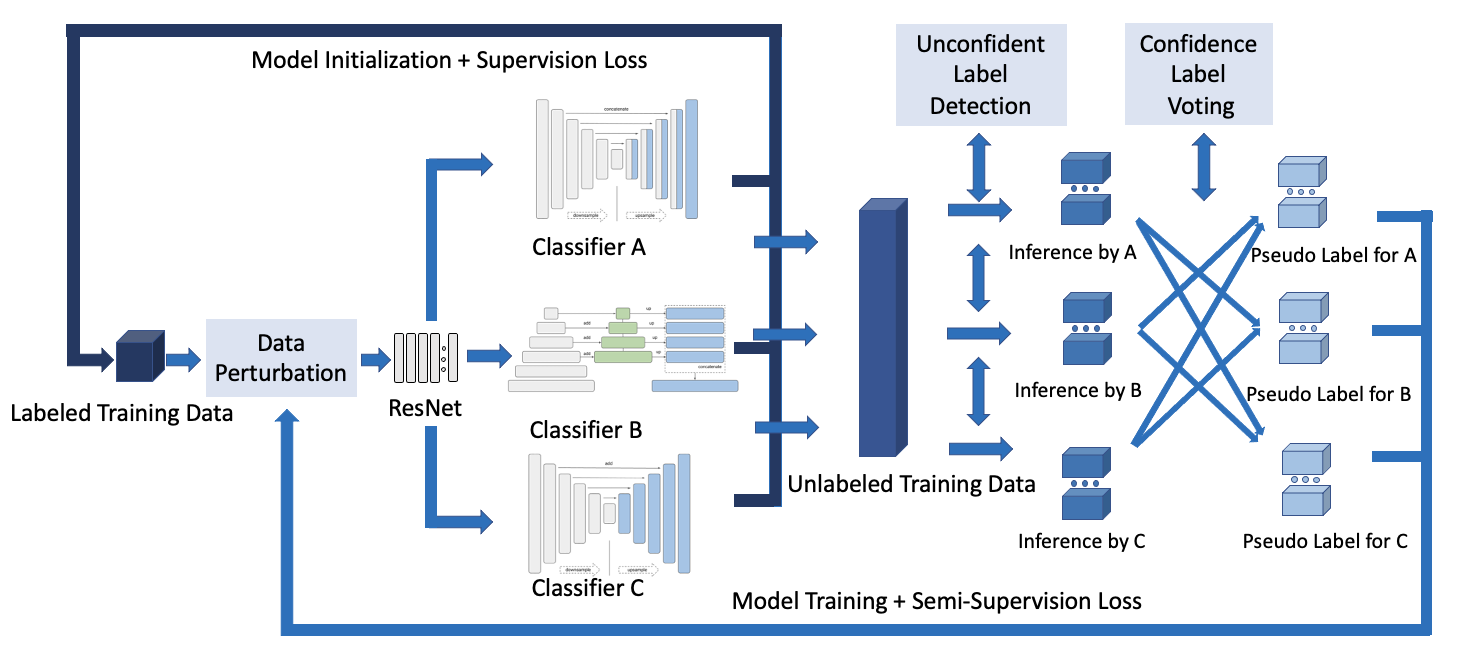}  
\caption{The Architecture of TriSegNet Framework}

\label{fig:trisegnet}  
\end{figure*}

The architecture of our {\bf Tri}ple-view feature learning for medical image semantic {\bf Seg}mentation {\bf Net}work (TriSegNet) is illustrated in Figure~\ref{fig:trisegnet}. 
%We discuss in some detail the training strategy that enables the framework to apply to the medical image segmentation domain. 
The label editing including unconfident label detection and confidence label voting based on a confidence estimate improves the feature expression of un-annotated data. Each of the three views of the feature learning makes use of random data perturbation for regularization. 
%We also propose a bespoke mixed boundary- and overlap-based loss function for the semi-supervised learning, applied once the confidence of the pseudo label generation is high enough.
%TriSegNet is tested on four open public datasets and a wide variety of evaluation measures to demonstrate competitive performance against other state-of-the-art semi-supervised algorithms. 

%In Figure~\ref{fig:trisegnet}, 
A pre-trained ResNet is utilized as a low-level feature learning module, shared between the three high-level feature learning pixel-level classifiers A, B and C. For the purpose of triple-view learning~\cite{chen2021semi}, the architecture and parameters initialization are developed separately. A, B, C are Encoder-Decoder-based.
In order to transfer sufficient semantic information, they fully model long-range dependencies, variation size of feature expression, model skip connection~\cite{ronneberger2015u}, they bypass spatial information~\cite{chaurasia2017linknet}, and process multi-scale feature maps~\cite{kim2018parallel}. Inspired by~\cite{donahue2014decaf}, the low-level feature learning module is shared, and the three views of classifiers enable the ResNet to extract low-level features in a generic manner. A, B and C not only extract features, but also vote, and generate pseudo labels, generally benefiting each other in the semi-supervised process.

\subsection{Training Setup}\label{Training}

%\begin{wrapfigure}{r}{5.8cm}
\begin{wrapfigure}{r}{0.48\textwidth}
\centering  
\includegraphics[width=0.45\textwidth]{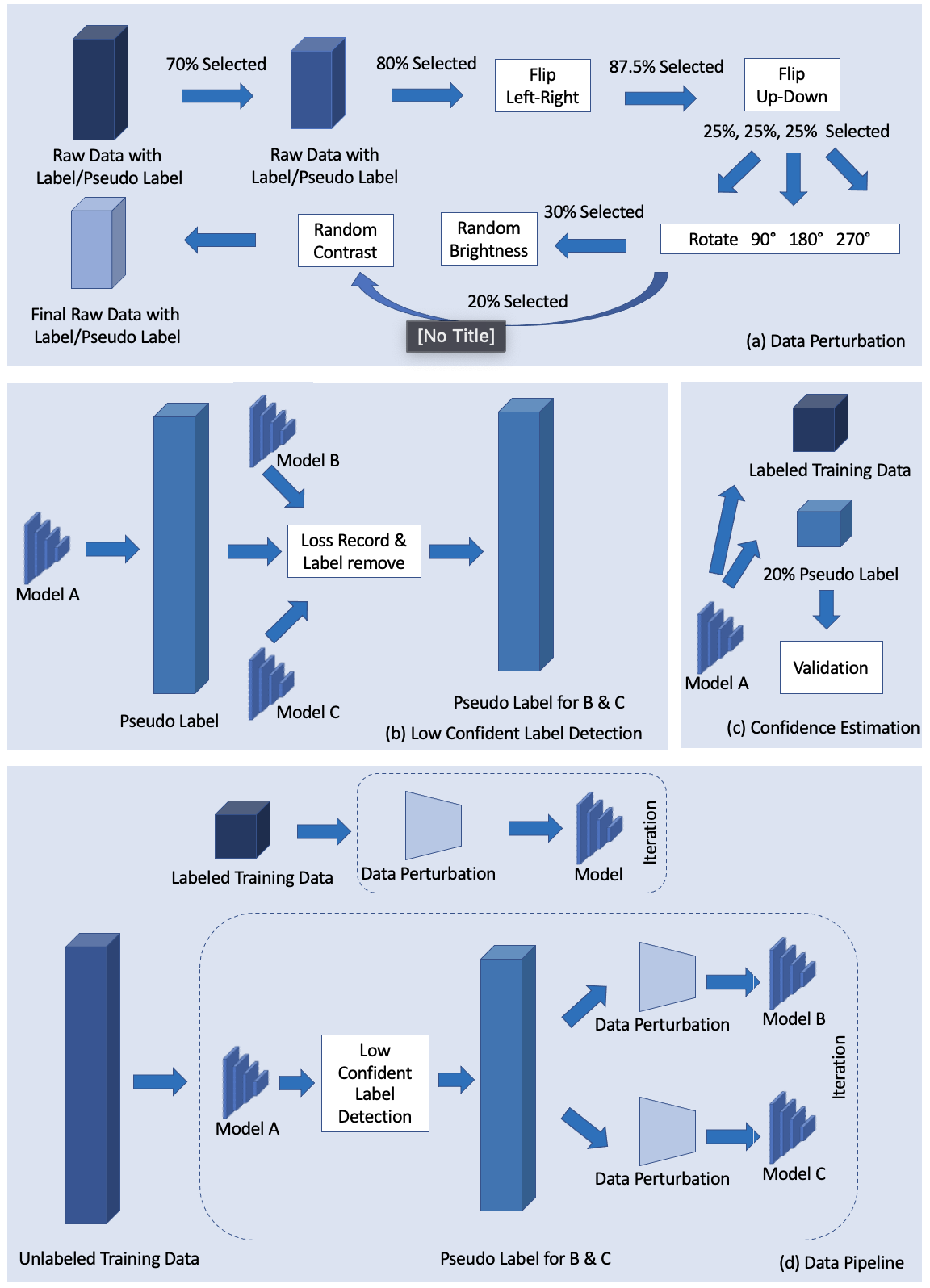}  
\caption{Data Pipeline and Label Processing stages}
\label{fig:da}  
\end{wrapfigure}

%In the task of semi-supervised learning, 
We denote by $\mathcal{L}$, $\mathcal{U}$, $\mathcal{T}$  a small labelled dataset, a large size of unlabelled dataset, and a dataset for testing. Each batch of labeled data is $(x, y_{gt}){\in}\mathcal{L}$, $(x, y_{gt}){\in}\mathcal{T}$; batches of only raw data are $(x){\in}\mathcal{U}$; $y_\mathit{pred}, y_\mathit{pseudo}$ is the dense map generated by different views of TriSegNet for prediction. We split the training process into three stages with an inference stage. 
{\bf  Stage 1}  initializes the three views of classifiers $f_{n}, n={1,2,3}$ with labeled dataset $\mathcal{L}$, repeatedly, 
as shown in Fig.~\ref{fig:da}(d), with data perturbation as per Fig.~\ref{fig:da}(a). To properly initialize the three classifiers with unbalanced data, overlap loss is used. The confidence weight during the training process is estimated and recorded as $\alpha_{n}, n={1,2,3}$ until inference is complete. Each model estimates its confidence of the labelled data by  on randomly selected pseudo labels, also compared against the other two models as sketched in Figure~\ref{fig:da}(c). 
{\bf  Stage 2}  learns from the generated pseudo labels in an iterative manner. Unlabeled data $(x)\in \mathcal{U}$ are used by $f_{n}$ as $(x, y_{n\_pred})$ pairs. Any data corresponding to low confidence labels gets removed.
%A specific portion of data are detected and removed as low confidence label, 
A fresh vote from another model $f_{n}$ generates pseudo label $(x, y_{n\_pseudo})=\alpha_{n-1}{\times}f_{n-1}(x) + \alpha_{n+1}{\times}f_{n+1}(x) $ to train the rest of the model. 
%Label processing including data perturbation, low confidence label detection and confidence label voting showing in Figure~\ref{fig:da}(b,c) are detailed discussed in Section~\ref{LabelProcessing}. 
The loss function at this stage is mixed overlap- and boundary-based loss. %to fine tune the model once triple classifiers can further confidently generate pseudo labels. 
{\bf  Stage 3} is to train additionally the lowest confident classifier with $\mathcal{L}$, $\mathcal{U}$. The learning rate for all training stages is set to $2{\times}10^{-4}$, with an Adam optimizer. The training epoch is set to 150 in Stage 1, 100 in Stage 2, and 150 or early stopped until the least confident classifier reaches the same confidence as other classifiers in Stage 3. The batch size is set to 16. {\bf Inference} is to predict jointly on an unseen image $(x)\in \mathcal{T}$ to $y_{pred}=\sum_{n=1}^{3} \alpha_{n}{\times}f_{n}(x) $.

\subsection{Label Processing}\label{LabelProcessing}

To encourage meaningful differences among views for co-training, it is essential to pin down each classifier's architecture, as well as its label editing. 
%Unlike with injecting random noise into true labels and generates modules from the diverse training~\cite{breiman2000randomizing}, or roughly split one dataset to multi dataset, or collect 2 dimension medical data from 3 dimension medical data in different view, we
In TriSegNet we  propose three label editing approaches: Data perturbation, low confidence label detection, and Confidence label voting.
%as label processing for TriSegNet and discussed in the following sections.

% \subsubsection{Data Perturbation }
% \label{DataAugmentation}

{\bf Data Perturbation.} Mean Teacher~\cite{tarvainen2017mean} uses perturbation to assess the consistency of the same image under different disturbances. This is  extended in Dual Student to pixel-wise tasks in guided collaborative training~\cite{ke2020guided}.  Naive-student~\cite{chen2020naive} uses perturbation for pseudo labels in an iterative manner to train the student network. To effect differences between the three views, TriSegNet also uses perturbation in an iterative manner, in each training epoch and separately for each of the views, as detailed in Fig.~\ref{fig:da}. 
From the input data, 70\% is selected, then 80\% of this is processed by left-right and up-down flips, and 25\% has 90$^\circ$, 180$^\circ$, 270$^\circ$ rotations applied. Random brightness and contrast changes are applied to 20\% and 30\% of the data respectively. Data perturbation  is used both on labelled data and on the pseudo labels predicted from each classifier.
%which get fed back into the process. 

% \subsubsection{Low Confidence Label Editing } \label{LabelEditing}

{\bf Low Confidence Label Detection.} A badly performing view may negatively influence the whole framework in that it  predicts pseudo labels for the other two views. Uncertainty estimation might be tackled with agreement-based entropy-maps or using pseudo labels for training, or computing uncertainty-weighted with a Bayesian deep network by adding dropout layers~\cite{xia20203d,chen2021semi,wang2022uncertainty}. Instead, we mark any unreliable pseudo labels as noisy~\cite{huang2019o2u,wang2021rar}, thus  temporarily removing low confidence labels from one view with the help of the other two views.  Retaining only high confidence pseudo labels boosts the overall performance. A simple adaptive denoising learning strategy is to calculate the overlap-based difference between each prediction in different views and record it for each pseudo label in each training epoch. The higher the label difference, the higher the probability of it not being accepted. During each training epoch with prediction label $(x,y_\mathit{pred})$, our strategy detects and removes specific number of the high difference labels $(x,y_\mathit{pseudo})$ raised by the other two views. More low confidence labels get detected at the beginning of the training iteration, and fewer towards the end., because the training process evolves from underfitting to overfitting. 
The number $N(t)$ of removed labels is

\begin{equation}\label{cleannoisylabel}
N(t) = \left\{
            \begin{array}{lr}
             0.05 (1-\zeta) y , & 0 < t <0.01 (1-\zeta) x \\
             \frac{-y}{x}t + 0.06(1-\zeta) y, & ~~~~~~0.01 (1-\zeta) x \leq t\& t \leq 0.05 (1-\zeta) x\\
             0.01 (1-\zeta) y , &  0.05 (1-\zeta) x < t \leq x
             \end{array}
\right.
\end{equation}
where $t$ is the current training iteration, $\zeta$ is the difference/disagreement level, $x$ is the total number of training iterations, and $y$ is the total number of $y_\mathit{pred}$.

% \subsubsection{Confident Label Voting}

{\bf Confidence Label Voting.} The uncertainty of the pseudo labels can sabotage expanding the training data, potentially influencing the whole framework performance. Luo~\cite{luo2021efficient} studied pyramid prediction network and uncertainty rectified pyramid consistency. Yu~\cite{yu2019uncertainty} proposed a Monte Carlo Dropout to generate an uncertainty map in student-teacher networks. Xia~\cite{xia20203d} proposed uncertainty weighted label fusion. Unlike the student-teacher network or single encoder-decoder networks, or directly calculating the average maximum posterior probability of another two views, Tri-net~\cite{dong2018tri} proposes a direct confidence estimate. In our algorithm, labelled training data and 20\% of un-perturbed pseudo label data is considered as a validation set for each view to obtain confidence weights; thus two views jointly generate pseudo labels for the last view using confidence voting together after low confidence label detection for cross pseudo supervision~\cite{chen2021semi}.

\subsection{Loss Function}\label{LossFunction}

% ~\cite{chen2019learning}~\cite{chan1999active}~\cite{bresson2007fast}
% The choice of loss function is always crucial.
%Dice-coefficient and Cross-Entropy both are two most common utilized loss function for segmentation in machine learning community~\cite{wang2021quadruple}~\cite{wang2021rar}.  These two loss functions, however, are only based on pixel-wised difference between predicted and ground truth~\cite{yeghiazaryan2018family}. 
% Conventional loss functions based on cross-entropy only consider pixel-wise differences. In order to address differences in the geometry, the circumference and area of the region are considered simultaneously to design a novel loss function. Normally t

The weight of pseudo-label-based semi-supervision loss relative to labeled-based supervised loss increases gradually during training, as the model becomes progressively more confident in generating and using pseudo labels~\cite{chen2021semi,laine2016temporal,tarvainen2017mean}. A low to high confidence strategy has been explored in uncertainty-aware schemes~\cite{yu2019uncertainty}. 
%Unlike with previous setting, dual-stage loss is conducted to 
TriSegNet enables  coarse-to-fine model training using gradually more precise pseudo labels: in Stage $1$ an overlap-based loss is used (Equation~\ref{loss1}), whereas Stages $2$ and $3$ rely on a mixed overlap- and boundary-based loss (Equation~\ref{loss2}).
%The details of each loss function are illustrated in Equations~\ref{loss1} and~\ref{loss2}.
Overlap-based loss is more robust when the datasets are imbalanced (e.g.\ fewer True pixels).  Once the pseudo labels from each view become dependable, the dual loss  is triggered, enabling the three views to segment more precisely~\cite{chen2019learning}.
%~\cite{chen2020aceloss}

% (2) To tackle the data imbalance problem (a common challenge in medical image segmentation), constants $\alpha$ and $\beta$ in Equation~\ref{loss1} are utilized~\cite{salehi2017tversky} to adjust how harshly different types of error are penalised. (3) To control the non-linearity of the loss~\cite{lin2017focal}, $\gamma$ is utilized in Stage 1, aiming to focus the model to learn features on harder examples, especially small scale segmentations. (4)

\begin{equation}\label{loss1}\small
\mathit{Loss}_\mathit{stage1} = \left(1-\frac{TP +  10^{-6}}{TP+\alpha FN+\beta FP + 10^{-6}}\right)^{\gamma}
\end{equation}

% \begin{equation}\label{loss2}\small
% Loss_{stage2\&3} = Loss_{Length} + Loss_{Region}
% \end{equation}

\begin{equation}\label{loss2}\small
\begin{split}
Loss_{stage2\&3} = \underbrace{\sum_{i,j=1}^{256} \sqrt{\| \bigtriangledown P_{x_{i,j}}  + \bigtriangledown P_{y_{i,j}}  + 10^{-6} \| }}_{Loss_{Boundary}} + \\ \underbrace{\| \sum_{i,j=1}^{256} P_{i,j}(1 - G_{i,j})^2 \| +  \| \sum_{i,j=1}^{256} (1 - P_{i,j}) G_{i,j}^{2} \|}_{Loss_{Overlap}}
\end{split}
\end{equation}

The ground truth, and pseudo label are denoted as $G, P{\in}[0,1]^{256{\times}256}$, where 1 and 0 denote the region of interest or  the background on a $256{\times}256$ image.
Constants $\alpha$, $\beta$ and $\gamma$ control the non-linearity of the loss~\cite{lin2017focal}.

% More specifically, $Loss_{Length}$ and $Loss_{Region}$ are illustrated in Equations~\ref{Length}~and~\ref{Region}.

% \begin{equation}\label{Length}\small
% \mathit{Loss}_\mathit{Length} = \sum_{i,j=1}^{256} \sqrt{\| \bigtriangledown P_{x_{i,j}}  + \bigtriangledown P_{y_{i,j}}  + 10^{-6} \| }
% \end{equation}

% \begin{equation} \label{Region}\small
% \begin{split}
% \mathit{Loss}_\mathit{Region} = & \| \sum_{i,j=1}^{256} P_{i,j}(1 - G_{i,j})^2 \| +  \| \sum_{i,j=1}^{256} (1 - P_{i,j}) G_{i,j}^{2} \|
% \end{split}
% \end{equation}

%where $10^{-6}$ in Equation~\ref{Length} helps avoid the `Zero Square' issue, and $ 256$ indicates the image size in each dimension.

\section{Experiments and Results}

\subsection{Datasets and Experimental Setup}

Our experiments have been carried out on four public benchmark datasets: (1)~The Ultrasound Nerve Segmentation  Kaggle Challenge~\cite{nervedataset}    Identifying nerve structures in ultrasound images is critical to  inserting a patient’s pain management catheter. We use 5,635 images of size $580{\times}420$. 
(2)~The MRI Cardiac Segmentation is from the Automated Cardiac Diagnosis MICCAI Challenge 2017~\cite{bernard2018deep}.  We use 1,203  images of size $232{\times}256$, from 150 patients.
(3)~The Histology Nuclei Segmentation is a Pan-Cancer histology dataset from the University of Warwick~\cite{gamper2019pannuke}. We use 7,901  images of size $256{\times}256$.
(4)~The CT Spine set includes 10 CTs from a CSI 2014 Segmentation Challenge~\cite{yao2012detection}, covering the entire thoracic and lumbar spine. We use 5,602  images of size $512{\times}512$.
%Ultrasound Nerve~\cite{nervedataset}, CT Spine~\cite{yao2012detection}, MRI Cardiac~\cite{bernard2018deep}, and Histology Nuclei~\cite{gamper2019pannuke}.

Ultrasound, MRI and Histology images have been resized to $256{\times}256$, and CTs to $512{\times}512$. All four are pre-processed with data perturbation. The data has been split into labelled training data, test data and unlabeled raw training data. The test data is always 10\% of the set, selected randomly for each run. The training data is the remainder of the dataset; for separate experiments, 2\%, 5\%, 10\% or 20\% of this is available labelled, and the rest is available raw. The validation data is selected as a separate random 20\% sample of the training set (both labelled and not), and is never seen by the training process. 

We used  Python 3.8.8, Tensorflow 2.6.0~\cite{tensorflow2015-whitepaper} and CUDA 11.3, on four Nvidia GeForce RTX3090 GPUs. The runtimes averaged 2--3.5 hours on the MRI data, and 6--8 hours on the Ultrasound, CT and Histology data, including the data transfer, three training stages, inference and evaluation. 
Semi-supervised learning for medical image segmentation,\footnote{https://github.com/HiLab-git/SSL4MIS} an online collection of implementations has been used in the baseline testing models and additional Dropout.
%without further changing are developed with. 
% We report the mean (but omit variance) for each measure over ten independent runs.

% \setlength{\tabcolsep}{3pt}
\begin{table}[!htbp]
\small

% \centering
\begin{tabular}{ccc|cc|cc|cc}
\hline
\hline
\multicolumn{9}{c}{Experiments Under the Assumption of {\bf 2\%} Data as Labeled Data} \\
\hline

  \multirow{2}{0.6in} & \multicolumn{2}{c|}{Ultrasound Nerve} & \multicolumn{2}{c|}{CT Spine} & \multicolumn{2}{c|}{MRI Cardiac} & \multicolumn{2}{c}{Histology Nuclei} \\
 \cline{2-9}
 &  IOU  & Sen  & IOU  & Sen  & IOU  & Sen  & IOU  & Sen  \\

 \hline

 {UNet}    &0.1628  & 0.2020 &  0.8657   & 0.9210 & 0.3888  & 0.8351 & 0.6574  & 0.7814\\
                    % & Variance   & 0.0096& {\bf 1.07e-7} & 0.0188 & 0.0013 & 0.30e-7 & 0.0011 & 0.0082 & 0.0006 & 0.0210 & 0.0037 & 0.0002 & 0.0047\\
  \hline
%  \multirow{2}{0.6in}
 {Linknet}        & 0.0919  & 0.1280 & 0.8438  & 0.8962 & 0.1498 & {\bf 0.9329 } & 0.6905 &  0.8219\\
                    % & Variance   & 0.0045 &3.07e-5 & 0.0127 & 0.0025 & 0.44e-7 & 0.0026 & 0.0042 & 0.0176 & 0.0072 & 0.0002 & 0.59e-7 & 0.0004\\
 \hline
%  \multirow{2}{0.6in}
 {FPN}        & 0.1227  & 0.1320 & 0.8653  & 0.8990 & 0.4802  & 0.5143 & 0.6942 &  0.8284 \\
                    % & Variance   & 0.0046 & 2.77e-4 & 0.0059 & 0.0024 & 0.33e-7 & 0.0033 & 0.0373 & 0.0001 & 0.0452 & 0.45e-5 & 0.20e-5 & 12.1e-7 \\
 \hline
%  \multirow{2}{0.6in}
 {{\bf TriSegNet}}        & {\bf 0.2800}  & {\bf 0.3678} & {\bf0.9526} & {\bf 0.7024 } & {\bf 0.9923} & 0.8411 & {\bf0.6946} & {\bf 0.8293} \\
                    % & Variance   & {\bf 0.0036} & 5.55e-7 & {\bf 0.0097 } & {\bf 0.75e-7} & {\bf 0.10e-7} & {\bf3.14e-5} & {\bf 0.0008} & {\bf 17.9e-7 } & {\bf 0.0020 } & {\bf1.47e-7} & 0{\bf.38e-7} & {\bf0.38e-7}  \\

\hline
\hline
% \end{tabular}
% \caption{Direct Comparison Against Existing Fully-Supervised Algorithms under 2\% Labeled Data}
% \label{table2}
% \end{table*}

\multicolumn{9}{c}{Experiments Under the Assumption of {\bf 5\%} Data as Labeled Data} \\

% \begin{table*}[!htbp]

% \centering
% \begin{tabular}{c|ccc|ccc|ccc|ccc}
\hline
%  &  \multicolumn{3}{c}{Ultrasound Nerve} & \multicolumn{3}{c}{CT Spine} & \multicolumn{3}{c}{MRI Cardiac} & \multicolumn{3}{c}{Histology Nuclei} \\ 
% \hline
%  &  IOU & Spe & Sen  & IOU & Spe & Sen  & IOU & Spe & Sen  & IOU & Spe & Sen  \\
%  \hline
%  \multirow{2}{0.6in}
 {UNet}        & 0.2762  & 0.3234 & 0.9207& 0.9576 & 0.6805 & 0.8053 & 0.7075 & 0.5953 \\
                    % & Variance   & 0.0036 & 2.10e-7 & 0.0075 & 0.0002 & 3.80e-7 & 0.0001 & 0.0123 & 3.75e-5 & 0.0258 & 0.0016 & 1.63e-5 & 0.0044 \\
  \hline
%  \multirow{2}{0.6in}
 {Linknet}        & 0.2505 & 0.2885 & 0.9020 & 0.9519 & 0.6762 & 0.7988 & 0.7552  & 0.8645 \\
                    % & Variance   & 0.0022 & 0.75e-7 & 0.0048 & 0.0007 & 0.15e-7 & 0.0004 & 0.0045 & 0.68e-5 & 0.0118 & 0.0002 & 0.78e-5 & 0.0001 \\
 \hline
%  \multirow{2}{0.6in}
 {FPN}        & 0.2703  & 0.3093 &  0.9221  & 0.9521 & 0.7721  & 0.8351 & 0.7031 & 0.8333 \\
                    % & Variance   & 0.0018 & 1.34e-7 & 0.0036 & 2.48e-5 & 0.34e-7 & 0.0001 & 0.0007 & 0.12e-5 & 0.0026 & 0.0001 & 0.29e-7 & 2.1497 \\
 \hline
%  \multirow{2}{0.6in}
 {{\bf TriSegNet}}        & {\bf 0.3765} & {\bf 0.5090 } & {\bf0.9385} & {\bf0.9660} & {\bf 0.8094} &  {\bf 0.8880} & {\bf 0.7691} &  {\bf 0.8757} \\
                    % & Variance   & {\bf 0.0002} & {\bf 0.68e-7} & {\bf 0.0012} & {\bf0.17e-5} & {\bf 0.01e-7} & {\bf0.90e-5} & {\bf 0.0001} & {\bf 2.49e-7} & {\bf 0.0001} & {\bf 0.0001} & {\bf0.69e-7} & {\bf4.06e-5} \\

\hline
\hline
% \end{tabular}
% \caption{Direct Comparison Against Existing Fully-Supervised Algorithms under 5\% Labeled Data}
% \label{table5}
% \end{table*}

\multicolumn{9}{c}{Experiments Under the Assumption of {\bf 10\%} Data as Labeled Data} \\

% \begin{table*}[!htbp]

% \centering
% \begin{tabular}{c|ccc|ccc|ccc|ccc}
% \hline
%  &  \multicolumn{3}{c}{Ultrasound Nerve} & \multicolumn{3}{c}{CT Spine} & \multicolumn{3}{c}{MRI Cardiac} & \multicolumn{3}{c}{Histology Nuclei} \\ 
\hline
%  &  IOU & Spe & Sen  & IOU & Spe & Sen  & IOU & Spe & Sen  & IOU & Spe & Sen  \\
%  \hline
%  \multirow{2}{0.6in}
 {UNet}        & 0.3554  & 0.4090 & 0.9320  & 0.9660 & 0.8492  & {\bf 0.9253} & 0.8012  & 0.8880 \\
                    % & Variance   & 0.0012 & 2.10e-7 & {\bf 0.0024} & 0.0002 & 3.47e-7 & 0.0002 & 0.0006 & 0.22e-5 & 0.0007 & 0.0001 & 0.37e-5 & 0.0001 \\
  \hline
%  \multirow{2}{0.6in}
 {Linknet}        & 0.3464 & 0.3991 & 0.9247  & 0.9527 & 0.7832  & 0.8641 & 0.7957  & 0.8862\\
                    % & Variance   & 0.0007 & 0.75e-7 & 0.0012 & 0.0008 & 0.14e-7 & 0.0004 & 0.0016 & 0.38e-5 & 0.0026 & 3.25e-5 & 9.11e-7 & 0.0001 \\
 \hline
%  \multirow{2}{0.6in}
 {FPN}        & 0.2416 & 0.4866 & 0.8316  & 0.9721 & 0.8078  & 0.8591 & 0.8034 & 0.8973 \\
                    % & Variance   & 0.0147 & 0.0004 & 0.0366 & 0.0415 & 0.0011 & 0.0003 & 0.0039 & 0.69e-5 & 0.0060 & 0.0001 & 0.41e-5 & 2.79e-5 \\
 \hline
%  \multirow{2}{0.6in}
 {{\bf TriSegNet}}        & {\bf 0.4260}  & {\bf 0.5789} & {\bf0.9463}  & {\bf0.9714} & {\bf 0.8545}  & 0.9159 & {\bf 0.8114 }  & {\bf 0.8981} \\
                    % & Variance   & {\bf 0.0005} & {\bf 0.20e-7} & 0.0032 & {\bf4.81e-7} & {\bf0.01e-7} & {\bf0.86e-5} & {\bf 5.94e-5} & {\bf 1.32e-7} & {\bf 4.78e-5} & 0.0001 & {\bf 0.71e-7}  & 4.47e-5 \\

\hline
\hline
% \end{tabular}
% \caption{Direct Comparison Against Fully-Supervised Existing Algorithms under 10\% Labeled Data}
% \label{table10}
% \end{table*}

\multicolumn{9}{c}{Experiments Under the Assumption of {\bf 20\%} Data as Labeled Data} \\

% \begin{table*}[!htbp]

% \centering
% \begin{tabular}{c|ccc|ccc|ccc|ccc}
\hline
%  &  \multicolumn{3}{c}{Ultrasound Nerve} & \multicolumn{3}{c}{CT Spine} & \multicolumn{3}{c}{MRI Cardiac} & \multicolumn{3}{c}{Histology Nuclei} \\ 
% \hline
%  &  IOU & Spe & Sen  & IOU & Spe & Sen  & IOU & Spe & Sen  & IOU & Spe & Sen  \\
%  \hline
%  \multirow{2}{0.6in}
 {UNet}        & 0.4352  & 0.5254 & {\bf 0.9447 }  & {\bf 0.9699 } & 0.8984  & 0.9448 & 0.8104 &  0.9019 \\
                    % & Variance   & 0.0004 & 2.59e-7 & 0.0011 & 5.65e-5 & 0.86e-7 & 7.97e-5 & 1.95e-5 & 0.36e-7 & 6.05e-5 & {\bf 4.34e-5} & 0.15e-5 & 5.79e-5 \\
  \hline
%  \multirow{2}{0.6in}
 {Linknet}        & 0.4333  & 0.5237 & 0.9295 & 0.9678 & 0.8712  & 0.9258 & 0.8027  & 0.8971 \\
                    % & Variance   & 0.0007 & {\bf 1.16e-7} & 0.0014 & 0.0021 & 0.51e-7 & 4.52e-5 & 0.0001 & 1.86e-7 & 0.0003 & 0.0002 & 0.87e-5 & 0.0001 \\
 \hline
%  \multirow{2}{0.6in}
 {FPN}        & 0.3956  & 0.5953 & 0.7605 & 0.8920 & 0.8857  & 0.9307 & 0.8176 & 0.9094 \\
                    % & Variance   & 0.0016 & 0.59e-5 & 0.0044 & 0.0928 & 0.0082 & 0.0450 & 7.38e-5 & 1.12e-7 & 0.0003 & 4.56e-5 & 0.17e-5 & 1.04e-5 \\
 \hline
%  \multirow{2}{0.6in}
 {{\bf TriSegNet }}        & {\bf 0.4981}  & {\bf 0.6528} & 0.9337  & 0.9494 & {\bf 0.9020 }  & {\bf 0.9459 } & {\bf 0.8530 } & {\bf 0.9244} \\
                    % & Variance   & {\bf 0.0003} & 4.49e-7 & {\bf 0.0006} & {\bf 0.25e-5} & {\bf 0.01e-7} & {\bf 1.06e-5} & {\bf 0.78e-7 } & {\bf 0.15e-7 } & {\bf 1.71e-5 } & 0.0001 & {\bf 2.56e-7} & {\bf 0.38e-7} \\

\hline
\hline
\end{tabular}

\setlength{\belowcaptionskip}{-10pt}

\caption{Evaluation Results on Direct Comparison between TriSegNet and Existing Fully-Supervised Algorithms}
\label{table20}

\end{table}

\begin{figure*}[t]
% \centering  
\includegraphics[width=\linewidth]{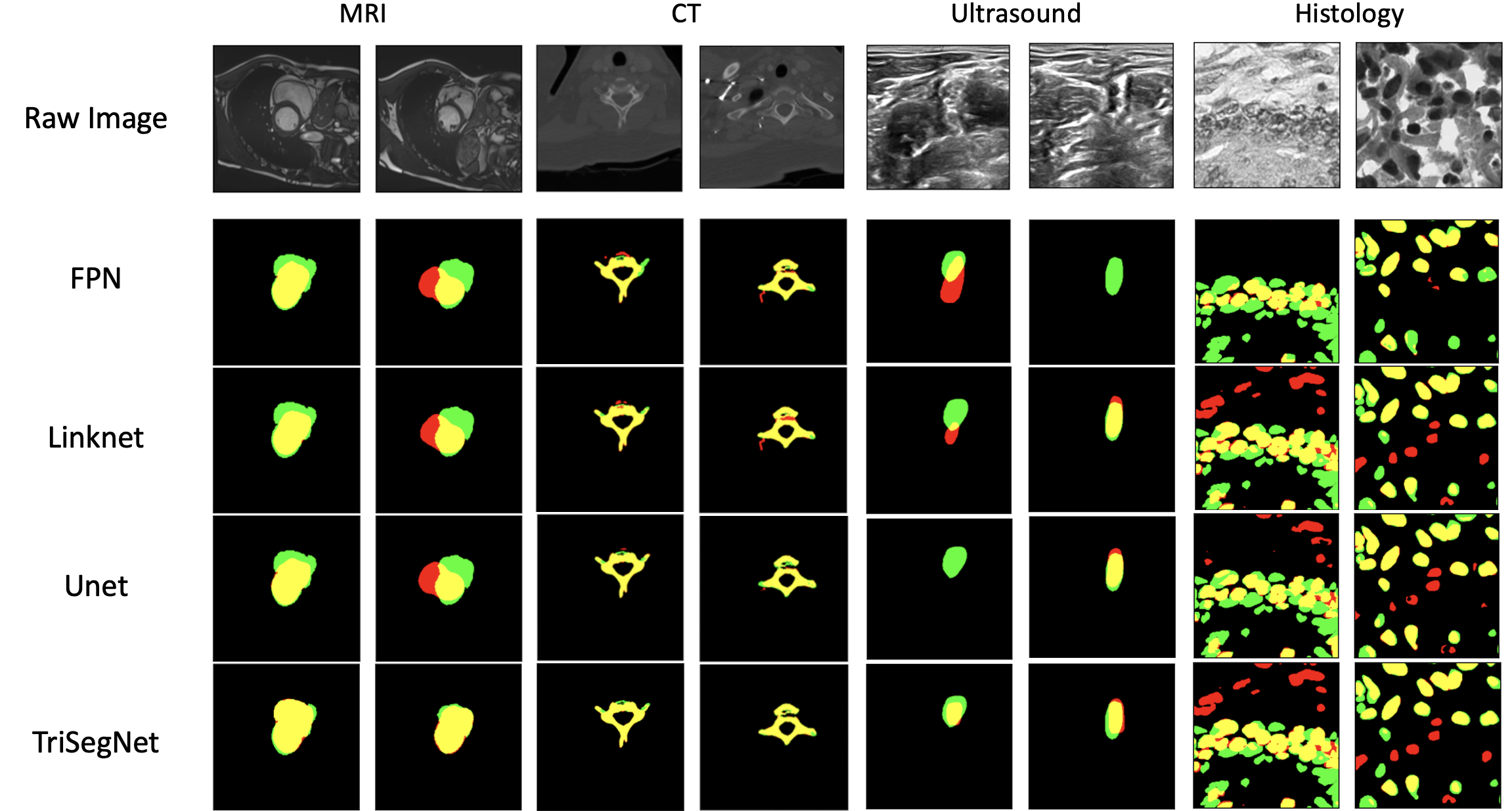}  
\caption{Sample qualitative results; Two images are selected from each test set where the first row illustrates the raw images. The rest of them illustrate the MS against GT where yellow, green, red, and black represent TP, FN, FP and TN at the pixel level.}
%Example results by TriSegNet and other fully-supervised algorithms}
\label{fig:trisegnetresultsv2}  
\end{figure*}

\subsection{Evaluation and Results}

Performance has been assessed using a wide range of evaluation metrics. These include the frequently reported similarity measures: Dice, IOU, accuracy, precision, sensitivity, specificity, as well as difference measures: relative volume difference (RVD), Hausdorff distance (HD), average symmetric surface distance (ASSD). To penalise mislabelled areas and avoid overly inflated segmentation scores, we also evaluate the boundary match between the machine segmentation (MS) and the Ground Truth (GT), using the Directed Boundary Dice relative to GT (DBD$_G$) and to MS (DBD$_M$) and Symmetric Boundary Dice (SBD)~\cite{yeghiazaryan2018family}. Experiments first compare TriSegNet with fully-supervised baseline methods like UNet\cite{ronneberger2015u}, LinkNet\cite{chaurasia2017linknet} and FPN\cite{kim2018parallel}. Table~\ref{table20} reports the quantitative results for each dataset, using 2\%--20\% of labeled data. The examples in Figure~\ref{fig:trisegnetresultsv2} illustrate these comparisons qualitatively.

\begin{table*}[!htbp]

% \centering
\begin{tabular}{cccccccccccc}
\hline
Model & Dice & Acc & Pre & Rec/Sen & Spe & RVD & HD & ASSD  & DBD$_G$ & DBD$_M$  & SBD \\
\hline
\cite{tarvainen2017mean} & 0.895  & 0.992 & 0.909 & 0.881 & 0.996 & 0.226  & 21.876 & 4.077 & 0.561 & 0.649 & 0.599 \\
\cite{vu2019advent} & 0.881  & 0.991 & 0.918 & 0.847 & 0.997 &  0.273  & 18.335 &  3.954 & 0.543 & {\bf  0.651} & 0.589 \\
\cite{zhang2017deep} & 0.878  & 0.990 & 0.858 & 0.899 & 0.994 & 0.332  & 24.235 & 5.046 & 0.577 & 0.564 & 0.559 \\
\cite{yu2019uncertainty} & 0.890  & 0.991 & 0.903 & 0.878 & 0.996 & 0.244  & 22.851 & 4.669 & 0.550 & 0.636 & 0.585 \\
\cite{verma2019interpolation} & 0.899  & 0.992 & 0.909 & 0.889 & 0.996 & {\bf 0.205}  & 28.388 & 4.773 & 0.582 & 0.639 &  0.605\\
% TriSegNet & 0.9165 & 0.9935 & 0.9091 & 0.9239 & 0.9963 & 0.3309  & 29.8705 & 5.6399 & 0.7024 & 0.3649 & 0.5185\\ % non augmentation

TriSegNet & {\bf 0.932} & {\bf 0.995} & {\bf 0.934} & {\bf 0.930 } & {\bf  0.997} & 0.208 & {\bf 7.831} & {\bf 2.075} & {\bf 0.712} & 0.611 & {\bf 0.657}\\
\hline
\end{tabular}
\setlength{\belowcaptionskip}{-10pt}

\caption{Direct comparison of TriSegNet with other algorithms, on the MRI Test Set}

\label{tablebaseline}
\end{table*}

To illustrate a wider range of evaluation measures with semi-supervised algorithms, there is only room to report on one dataset. The MRI Cardiac data has been chosen for this purpose. Table~\ref{tablebaseline} documents the performance of TriSegNet  against  Tarvainen~\cite{tarvainen2017mean}, Vu~\cite{vu2019advent}, Zhang~\cite{zhang2017deep}, Yu~\cite{yu2019uncertainty}, Verma~\cite{verma2019interpolation} with UNet as backbone, showing that it outperforms previous methods under most of the considered metrics.

%Table~\ref{tab:ablationcomparison} 
%to analyze the effects of each of the label processing contribution, dual loss design, the three different CNN architecture of classifiers as high-level feature learning views and their combinations and the suitability of proposing the different architecture of classifiers, extensive ablation experiments have been conducted. 

In order to assess the contribution of each of the components, specifically focused ablation experiments have been designed. They illustrate the essential role of models A, B and C being different by considering, instead of the three, two copies of one of them ($A{\times}2$,  $A{\times}3$, etc.). Table~\ref{tab:ablationcomparison} illustrates that the  IOU metric is negatively influenced by such choices, performing best when each of A, B and C is present in its own right.

\begin{table}[t]
\small

\caption{Ablation Studies on Contributions of Architecture and Modules}
\label{tab:ablationcomparison}
\begin{center}
% \centering
\begin{tabular}{ c|c|c|c|c|c }
\hline
\multirow{2}{3.3em}{Label Process} & \multirow{2}{4.4em}{Dual Loss Design} & \multirow{2}{4em}{Classifier A} & \multirow{2}{4em}{Classifier B} &  \multirow{2}{4em}{Classifier C} & \multirow{2}{3.5em}{IOU}\\
 & &  &  & & \\
 \hline
 
      &  & \checkmark $\times$ 2 & \checkmark &  & 0.8724 \\
      &  &  & \checkmark $\times$ 2 & \checkmark  & 0.8739 \\
      &  & \checkmark &  & \checkmark  $\times$ 2 & 0.8641 \\
    %  &  & \checkmark $\times$ 3 &  &  & 0.8652 \\
  \checkmark   &  & \checkmark $\times$ 3 &  &  & 0.8666\\
     & \checkmark  & \checkmark $\times$ 3 &  &  & 0.8579\\
     &  &  & \checkmark $\times$ 3 &  & 0.8598\\
       \checkmark   &  &   & \checkmark $\times$ 3&  & 0.8605\\
    %  & \checkmark  &  & \checkmark $\times$ 3 &  & \\
         &  &  &  & \checkmark $\times$ 3 & 0.8619\\
       \checkmark   &  &   & & \checkmark $\times$ 3 & 0.8739\\
    %  & \checkmark  &  &  & \checkmark $\times$ 3 & \\
    
 \checkmark  &  & \checkmark & \checkmark & \checkmark & 0.8787 \\
  &  \checkmark & \checkmark & \checkmark & \checkmark & 0.8841\\
\checkmark & \checkmark & \checkmark & \checkmark & \checkmark & 0.9020\\
 \hline

\end{tabular}
 \setlength{\belowcaptionskip}{-10pt}
\end{center}
\end{table}

\section{Conclusion}
Four medical datasets with different labelled/unlabelled assumptions have been used in the experiments. A series of experiments are designed including the comparison between TriSegNet and fully-supervised learning algorithms, as well as its comparison with other semi-supervised learning algorithms. Ablation studies justify the design decisions.
Although a common Dice-based loss function is used for the initial stages of the training process, a bespoke boundary-overlap-based loss is used in the more advanced stages. This increases the confidence of the model in its predictions and hence the reliability of the pseudo labels it generates.
Overall, TriSegNet demonstrates promising performance in most evaluation metrics, showing great potential in semi-supervised learning for general medical image segmentation.

\newpage

%%%%%%%%% REFERENCES
{
\bibliographystyle{splncs04}
\bibliography{egbib}
}

\clearpage
\appendix
% \begin{subappendices}

\section{Algorithm of TriSegNet}

The training of TriSegNet consists of four stages which is briefly illustrated in Algorithm~\ref{trisegnetalgorithm}. The code of TriSegNet will be publicly available.\footnote{https://github.com/ziyangwang007/CV-SSL-MIS}

%\begin{algorithm}[]
\begin{algorithm}
\SetAlgoLined
% \KwResult{Write here the result }
% \KwIn{
{\bf Input:} 
A batch of $(x, y)$ from labeled dataset $\mathcal{L}$, unlabeled dataset $\mathcal{U}$, or test dataset $\mathcal{T}$. 
$\mathit{DA}$, $\mathit{LCLE}$, and $\mathit{CLV}$ are label processing approaches Data Augmentation, Low Confidence Label Editing, and Confident Label Voting, respectively. \\
% }
{\bf Output:} 
Three trained high-level feature learning classifiers. $f_{n}, n={1,2,3}$

 \textbf{Stage one:} Initialization \\
 $\theta_{1}, \theta_{2}, \theta_{3} \leftarrow $ initial parameters of classifiers $ f_{1}, f_{2}, f_{3}$, and the loss function $L_{Supervision}$. \\
 
\While{ n $\leftarrow$ [1, 2, 3] (3 classifiers)}
{
$(x, y_{gt})$ sampled from $ DA(\mathcal{L}) $ with augmentation\;
Generate prediction $y_{pred} = f_{n}(x)$  \;
Calculate loss $L_{Supervision}$ with $y_{pred} = y_{gt}$ \;
Update $\theta_{n} \leftarrow \theta_{n} - \Delta L $ \;
}

\textbf{Stage two:} Classifiers Training with Pseudo Label Processing\\
$\alpha_{1}, \alpha_{2}, \alpha_{3} \leftarrow $ initial confidence weight of classifiers $ f_{1}, f_{2}, f_{3}$ \;
\While{ n $\leftarrow$ [1, 2, 3], i $\leftarrow$ [1...5] (5 Iterations)}
{
$x$ sampled from $DA(\mathcal{U})$ with augmentation \;
Generate pseudo label for Classifier $ f_{n} $ $y_{n\_pseudo} = CLV( LCLE(f_{n-1}(x),  LCLE(f_{n+1}(x))) $ \;
Calculate loss $L_{SemiSupervision}$ with $y_{pred} = y_{n\_pseudo}$ \;
Update $\theta_{n} \leftarrow \theta_{n} - \Delta L $ \;
Update $\alpha_{n} \leftarrow $ with $L$ by evaluation dataset \;

}

\textbf{Stage three:} One low confidence Classifer Training \\
\While{ n $\leftarrow$ [1, 2, 3], i $\leftarrow$ [1...5]}
{
{\bf If} Network $ f_{n} $ is with the lowest $\alpha_{n}$  {\bf Then}

$(x, y_{gt})$ sampled from $ DA(\mathcal{L}) $ \;
$(x, y_{n\_pseudo})$ sampled from $DA(\mathcal{U}) $ \;
Generate pseudo label for Classifer $ f_{n} $ $y_{n\_pseudo} = CLV( LCLE(f_{n-1}(x),  LCLE(f_{n+1}(x)))  $ \;
Calculate loss $L$ with $y_{pred} = y_{n\_pseudo}$ \;
Update $\theta_{n} \leftarrow \theta_{n} - \Delta L_{SemiSupervision} $ \;
Update $\alpha_{n} \leftarrow $ with $L$ by evaluation dataset \;

}

\caption{Training Stage of TriSegNet for Medical Image Segmentation}
\label{trisegnetalgorithm}  
\end{algorithm}

% \end{subappendices}

\section{The CNN Architecture of Multi-View Learning}

To properly encourage the differences of the three views of feature learning on dense prediction, not only the data feed and initialization of parameters, but three different advanced CNN are proposed in TriSegNet. We utilize three different techniques for CNN i.e. skip connection, efficiently passing feature information through residual learning, and multi-scale feature learning. The parameters of three classifiers are briefly illustrated in Table~\ref{tablemodel} and the source code has been released online.\footnote{https://github.com/qubvel/segmentation\_models/tree/master/segmentation\_models}

%\begin{table*}[]
\begin{table*}

\centering
\begin{tabular}{c|ccc}
\hline
Model & Total Params & Trainable params & Non-trainable params\\
\hline
Classifier A & 32,561,120 & 32,513,562 & 47,558 \\
Classifier B & 20,323,985 & 20,317,169 & 47,558 \\
Classifier C & 17,594,453 & 17,592,149 & 2,304  \\
\hline
\end{tabular}

\caption{The Computation Cost Information of Three Classifier}

\label{tablemodel}
\end{table*}

\section{Evaluation Methods, Qualitative, and Quantitative Results}

Table~\ref{tablebaseline} reports the TriSegNet performance direct comparison with other algorithms with several strict and novel quantitative evaluation metrics to which the boundaries of the machine segmentation(MS) match those of the ground truth(GT), using the Directed Boundary Dice relative to GT (DBD$_G$), Directed Boundary Dice relative to MS (DBD$_M$) and Symmetric Boundary Dice (SBD). 

In a von Neumann neighbourhood $N_x$ of each pixel $x$ on the boundary $\partial G$ of the ground truth,
\begin{equation}
\small
DBD_G=DBD(G, M)=\displaystyle\frac
    {\rule{0pt}{3ex}
    \sum\limits_{x \in \partial G} \text{Dice}(N_x)}
    {\left| \partial G \right|}
\end{equation}
\begin{equation}
\small
DBD_M=DBD(M, G)=\displaystyle\frac
    {\rule{0pt}{3ex}
    \sum\limits_{x \in \partial M} \text{Dice}(N_Y)}
    {\left| \partial M \right|}
\end{equation}
\begin{equation}
\small
    SBD =
    \displaystyle
    \frac
    {\rule{0pt}{3ex}
    \sum\limits_{x \in \partial G} DSC(N_x) + \sum\limits_{y \in \partial M} DSC(N_y)}
    {\left| \partial G \right| + \left| \partial M \right|}
\end{equation}
where Dice is $Dice(N_x) = \frac{2 | G(N_x) \cap M(N_y)|}{| G(N_x)| +  | M(N_y)|} $. The symmetric average is being brought down by DBD$_G$ when the latter features isolated areas of false negative labels. These measures penalise mislabelled areas in the machine segmentation.

\begin{figure*}[t]
\centering  
\includegraphics[width=\linewidth]{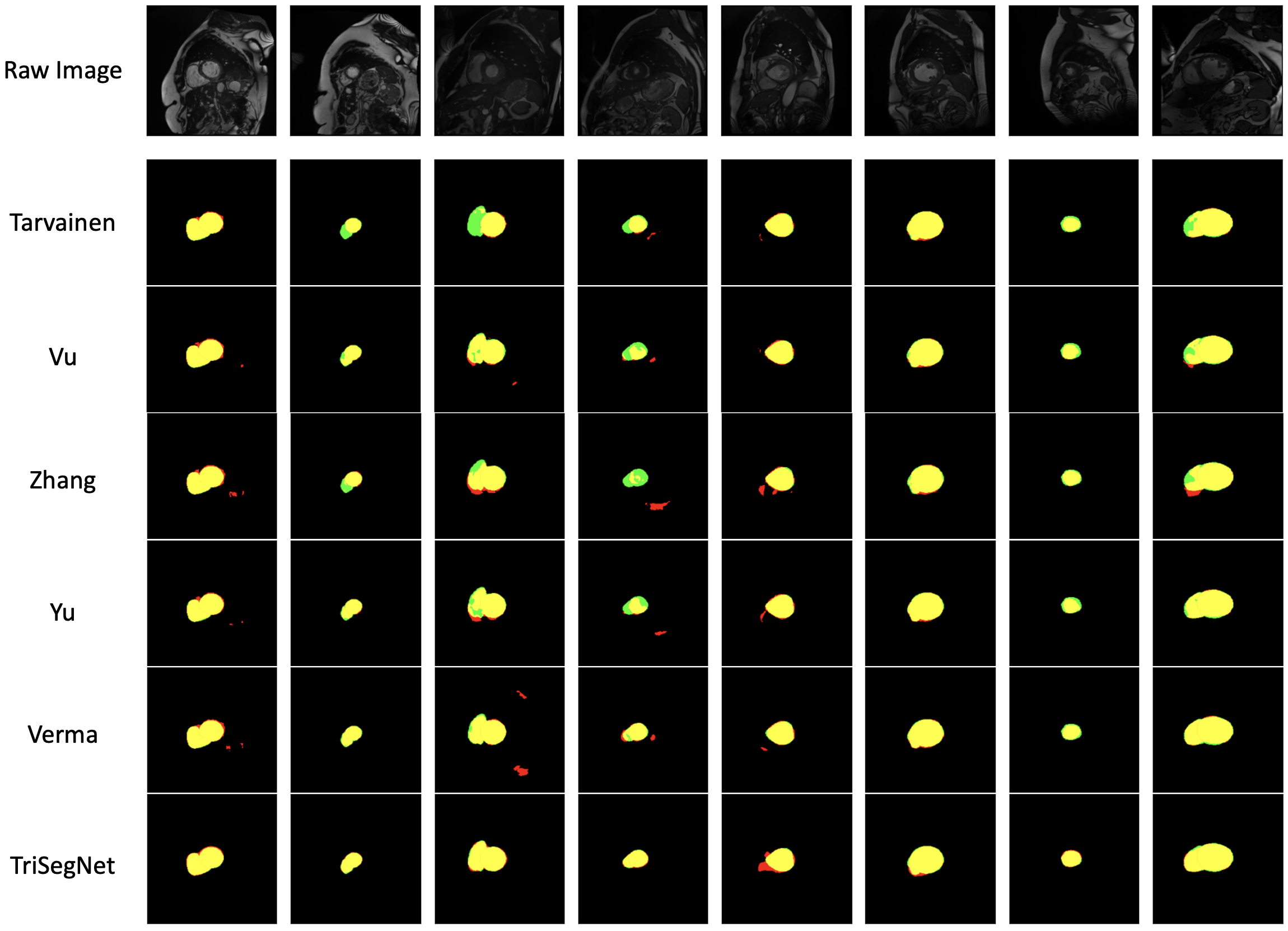}  
\caption{Sample Qualitative Results on MRI Cardiac Test Set}
\label{fig:trisegnetresultsv1}  
\end{figure*}

Some of example qualitative results on MRI Cardiac test set are briefly sketched in Figure~\ref{fig:trisegnetresultsv1}. Eight images are selected from MRI test set where the first row illustrates raw images. The rest of them illustrate the MS by each semi-supervised algorithm against GT where yellow, green, red, and black represent true positive, false negative, false positive and true negative at pixel level. The proposed method shows fewer false positive and false negative pixels, and significantly low HD as well, because the TriSegNet is beneficial with different views of high-level pixel-level classifier and proposed mixed boundary- and overlap-based loss function.

\end{document}